\documentclass[aps,prb,twocolumn,superscriptaddress]{revtex4-1}

\usepackage{graphicx}
\usepackage{feynmp}
\usepackage{epsfig}
\usepackage[latin1]{inputenc}
\usepackage{amsmath}
\usepackage{amsfonts}
\usepackage{amssymb}
\usepackage{amsmath}
\usepackage{amssymb}
\usepackage{color}
\usepackage{multirow,bigdelim}
\usepackage{enumerate}
\usepackage{amsthm}
\usepackage{amsbsy}
\usepackage{wasysym}
\usepackage{mathrsfs}
\usepackage{subfigure}

\newcommand{\tk}{t_{\scriptscriptstyle K}}
\newcommand{\lk}{l_{\scriptscriptstyle K}}
\newcommand{\zt}{$\mathbb{Z}_2 \, \,$}
\newcommand{\be}{\begin{equation} }
\newcommand{\ee}{\end{equation} }
\newcommand{\ba}{\begin{eqnarray} }
\newcommand{\ea}{\end{eqnarray} }
\newcommand{\n}{\nonumber \\ }

\newcommand{\mc}[1]{\mathcal{#1}}

\begin{document}

\title{Kibble-Zurek Scaling and String-Net Coarsening in Topologically Ordered Systems}

\author{Anushya Chandran}
\affiliation{Department of Physics, Princeton University, Princeton, New Jersey 08544, USA}
\author{F. J. Burnell}
\affiliation{All Souls College, Oxford, United Kingdom} %
\affiliation{Theoretical Physics, Oxford University, 1 Keble Road, Oxford, OX1 3NP, United Kingdom}
\author{Vedika Khemani}
\affiliation{Department of Physics, Princeton University, Princeton, New Jersey 08544, USA}
\author{S. L. Sondhi}
\affiliation{Department of Physics, Princeton University, Princeton, New Jersey 08544, USA}
\date{\today}

\begin{abstract}
We consider the non-equilibrium dynamics of topologically ordered systems driven across a continuous
phase transition into proximate phases with no, or reduced, topological order. This dynamics exhibits
scaling in the spirit of Kibble and Zurek but now {\it without} the presence of symmetry breaking and
a local order parameter. The late stages of the process are seen to exhibit a slow, coarsening dynamics for
the string-net that underlies the physics of the topological phase, a potentially interesting
signature of topological order. We illustrate these phenomena in the context of
particular phase transitions out of the abelian \zt topologically ordered phase of the toric code/\zt gauge theory,
and the non-abelian SU(2)$_k$ ordered phases of the relevant Levin-Wen models.
\end{abstract}
\maketitle

\section{Introduction}
This paper lies at the intersection of two interesting streams of contemporary research: the study of the non-equilibrium dynamics of quantum systems, and the study of topologically ordered phases of matter.
Consider driving a topologically ordered system through a phase transition to a topologically trivial, or relatively trivial, phase by changing some parameter in the Hamiltonian at a slow, but finite rate.
This work investigates the ensuing non-equilibrium dynamics resulting from this ``trans-critical protocol'' \cite{Chandran}, with particular emphasis on universality and the
non-Landau character of the transition.

The Kibble-Zurek (KZ) mechanism \cite{Kibble1976, Zurek1985, Zurek1996} is a scaling theory of the defects generated by slowly cooling a classical system through a continuous symmetry-breaking phase transition. Kibble originally formulated this problem as a cosmological theory of the phase transitions in an expanding universe, and Zurek later applied it to condensed matter systems. The central physical insight is that sufficiently distant parts of the system settle into independently seeded local broken symmetry ``directions'' whence the mismatch must be accommodated by a finite density of defects. The mechanism has since been generalized to quantum phase transitions \cite{Polkovnikov2005, Dziamarga2005, Zurek2005, Damski2005}. 
Although several experiments are consistent with the predictions of Kibble-Zurek \cite{Ducci, Maniv, Monaco}, decisive confirmation of the scaling law of defect densities is still lacking.
A recent experiment in an inhomogenous system of trapped ions \cite{KZexp} provides the most compelling evidence in this regard \footnote{Confirmation is still lacking in transitions in thermodynamic systems that are not described within mean field theory.}. Polkovnikov and co-workers have also studied the scaling theory of the excess heat density and the interplay between the ramp velocity and finite size \cite{Polkovnikov:2005zr, De-Grandi:2010aa}. For a recent review of the broader context of these developments in the study of non-equilibrium quantum phenomena, see Ref.~\onlinecite{Polkovnikov:2011ys} . 

A few observables \cite{De-Grandi:2010aa, Deng, Biroli, Zurek2010} other than the universal non-equilibrium defect density have been studied in the KZ problem, now defined more broadly as the non-equilibrium temporal evolution of a system in the vicinity of a critical point. Recently, Chandran et. al. \cite{Chandran} have systematized the universal content in the KZ problem in a scaling limit, and written non-equilibrium scaling functions for \emph{all} physical observables in this limit. The scaling functions describe the entire time history, and asymptote to equilibrium and coarsening scaling regimes in the appropriate limits. The universal content depends only on the pairing of the dynamical universality class of the critical point, and the particular protocol through parameter space. 

The KZ scaling theory also provides an elegant framework within which to investigate ramp dynamics of phase transitions beyond those of traditional symmetry breaking. These transitions could involve the destruction of the topological order of states of matter like spin liquids and the fractional quantum Hall phases. Topologically ordered phases are {\it not} locally distinguished by any order parameter, but are characterized by emergent gauge fields and fractionalized excitations. Their non-local structure makes them particularly robust to local perturbations and well-suited to perform ``topological'' quantum computation \cite{Nayak:2008aa}. Despite their robustness, a strong enough perturbation can drive a transition from a topologically ordered phase to a trivial or relatively trivial (i.e. one with a smaller gauge group) phase. For appropriately selected perturbations, this transition will be continuous.

This paper addresses the KZ problem when a system is driven from a topologically ordered phase to a proximate trivial or relatively trivial phase.
We do so for a class of topological phases that possess lattice realizations where the gauge degrees of freedom
are manifest: these are the toric code/lattice \zt gauge theory \cite{Kogut, Kitaev} and the string-net models of Levin and Wen \cite{LW} that realize doubled non-Abelian Chern-Simons theories.
By a combination of duality and perturbative arguments, we show that KZ scaling in the generalized sense of Ref.~\onlinecite{Chandran} holds for various observables even
though the canonical KZ signature of a density of topological defects is not meaningful.
We further provide strong arguments that the late time dynamics in the scaling regime exhibits a slow coarsening of the string-net that is condensed in the starting topologically ordered state. To our knowledge, this is the first treatment of a quantum coarsening regime in the dynamics of an isolated quantum system.
As the extended string-nets are central to the topological character of the starting phase \footnote{Sensitivity to the topology of the lattice manifold requires extended degrees of freedom like strings. The strings form nets as the phase is a liquid.}, their slow decay outside the phase is a (potential) signature of the physics of the parent phase.
The restriction to the scaling limit always brings simplification as particular gapped degrees of freedom are, at worst, dangerously irrelevant. That is, they do not affect the scaling regime but do alter the asymptotically long time behavior of the KZ process.

The most relevant precursor to our work is found in the cosmology literature in the papers of Rajantie and Hindmarsh \cite{Hindmarsh:2000uq, RAJANTIE:2002ff}. They studied the non-equilibrium dynamics of ramps through the finite temperature phase transition in the non-compact Abelian Higgs model; their protocol moves between the gapless Coulomb phase with gapped matter to the fully gapped Higgs phase.
However, there are three important differences.
First, their work involves finite temperature in an essential way. The zero temperature limit of their protocol would involve exciting the system in the gapless phase even before the transition is reached. Second, the non-compactness of their gauge field makes the physics of their Higgs phase qualitatively different from that of the compact gauge models considered by us. This difference is quite visible in our choice of observables.
Third, we work on the lattice in the ``electric flux'' representation, a natural choice in the condensed matter setting, while they work in the continuum with the vector potential. Thus our discussion of string-net coarsening has no analog in their formulation. In the condensed matter literature, ramps across topological transitions\cite{DeGottardi:2011qo} in (1+1)D and sudden quenches in one of our model systems, the perturbed toric code, have been studied before \cite{Rahmani:2010kl, Tsomokos:2009oq}. The sudden quenches are in a completely different limit from the slow ramps we study here as they inject a large amount of energy into the system. Finally, low temperature spin ice exhibits topological order in a classical limit \cite{Castelnovo:2012lh} and its dynamics following quenches is dominated by monopoles of the gauge field. However in this case, non-universal lattice effects turn out to dominate the long time behavior \cite{Castelnovo:2010fu} .

We turn now to the contents of the paper.
We begin in Sec.~\ref{Sec:PhaseDZ2} by reviewing the phase diagram of the \zt gauge theory coupled to matter and describing the transitions out of its topologically ordered phase. Readers literate in the canon of topological phases can skim this section for our notation.
Section~\ref{Sec:KZ1} describes the KZ ramp across the pure matter sector of the \zt theory which has a conventional symmetry-breaking transition. This section contains a new analysis of coarsening in the (2+1)D transverse field Ising model and summarizes our previous understanding of KZ.
In ~\ref{Sec:KZ2}, we discuss our results on the KZ scaling functions and string-net coarsening for ramps in the pure gauge sector of the theory, which has a confinement transition without a local order parameter.
We then generalize the scaling theory to a ramp across an arbitrary point on the critical line in the phase diagram in ~\ref{Sec:KZ3}.
We turn to generalizations of these results to phases with non-Abelian topological order in Sec.~\ref{Sec:LW}; specifically, we discuss a particular transition from the $SU(2)_k$ ordered phases. We end with a discussion of generalizations to other theories.

\section{Review of the phase diagram of the \zt Gauge Theory}
\label{Sec:PhaseDZ2}
The phase diagram of the $(d+1) $ dimensional \zt gauge theory with matter \cite{FradkinShenker} contains a topologically non-trivial (deconfined) phase and a topologically trivial (confined-Higgs) phase. The topological order in the deconfined phase is described by the BF theory \cite{Hansson:2004ye}. We work in $d=2$ for which the \zt theory is precisely (the topologically ordered) Kitaev's toric code with perturbations\cite{Kitaev}. We start by reviewing the key features and the excitation spectrum of the toric code. We then discuss two perturbations that drive a continuous transition to a topologically trivial phase.

\subsection{The perturbed toric code}
\begin{figure*}
\includegraphics[width = 17 cm]{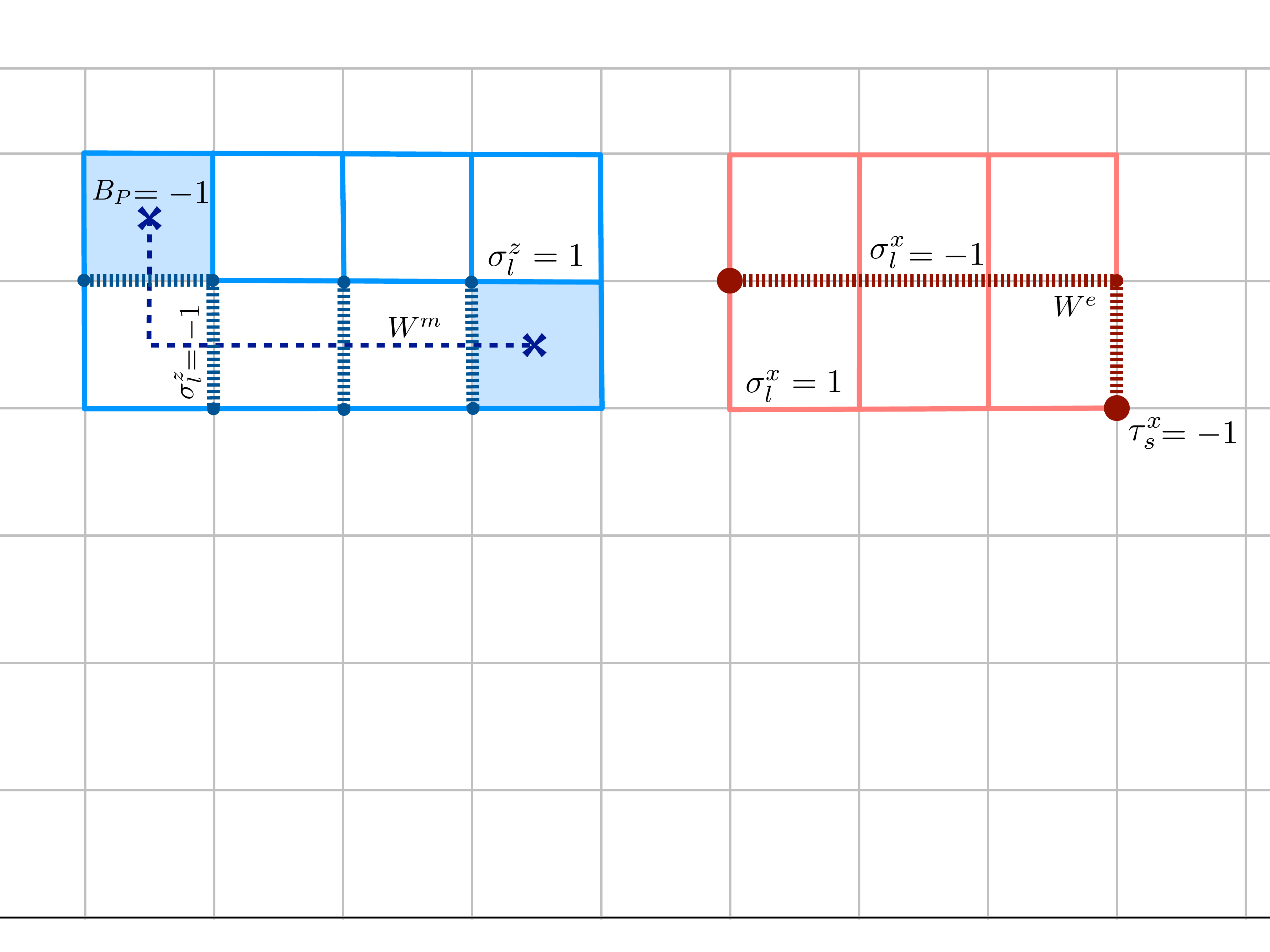}
\caption{A section of the toric-code lattice with operators in the magnetic (left) and electric (right) bases. Matter ($\tau$) and gauge ($\sigma$) variables are located on the sites and links respectively. The $W^m$ operator \eqref{Eq:Wm} flips a string of $\sigma^z$ variables and creates a pair of magnetic vortices at its endpoints, while the $W^e$ operator flips spins in the $x$ basis to create a pair of electric charges linked by electric flux.}
\label{Fig:TC}
\end{figure*}
The toric code\cite{Kitaev, Kitaev:2009qe} is defined in terms of spin-1/2 degrees of freedom that live on the links $l$ of a 2D square lattice:
\begin{align}
H_{\scriptscriptstyle TC} &= -K \sum_P B_P -\Gamma_M \sum_s A_s \nonumber\\
&\equiv  -K \sum_P\prod_{l\in \partial P} \sigma_l^z- \Gamma_M \sum_s \prod_{l: s \in \partial l } \sigma_l^x
\label{Eq:toric}
\end{align}
where the $A_s$ and $B_P$ are ``star'' and ``plaquette'' operators. $s$ and $P$ denote the sites and elementary plaquettes of the lattice, while $\partial P$ and $\partial l$ are the boundaries of plaquettes and links. $H_{\scriptscriptstyle TC}$ can be rewritten as a gauge theory with matter by identifying the $\sigma_l$ variables as the gauge degrees of freedom, and introducing new spin $1/2$ `matter' variables, $\tau_s$, on the sites of the lattice. Upon restricting the expanded Hilbert space to the `physical' subspace of gauge-invariant states
\begin{align}
G_s  |\psi\rangle = |\psi\rangle, \quad  \quad \quad  G_s = \tau_s^x \prod_{l: s\in \partial l} \sigma_l^x,
\label{Eq:Gaugeinvariant}
\end{align}
the toric code Hamiltonian (\ref{Eq:toric}) is equivalent to the gauge-invariant Hamiltonian:
\begin{align}
H_{0} =  -K \sum_P\prod_{l\in \partial P} \sigma_l^z- \Gamma_M \sum_s \tau_s^x .
\label{Eq:toric2}
\end{align}
Note that $G_s$ defines a set of local symmetries at each site since $[H_0, G_s] = 0 \; \forall \; s$.

In the $x$ basis of the spin operators, it is useful to think of  $\tau$ and $\sigma$ as the electric `charges' and `fluxes' in the theory respectively: $\tau_s^x = -1 \;(+1)$ if an electric charge is present (absent) at site $s$, while $\sigma_l^x = -1$ denotes the presence of electric flux on link $l$. In this language, we recognize the gauge-invariant condition (\ref{Eq:Gaugeinvariant}) as the lattice \zt version of Gauss's law. In the conjugate $z$ basis, the operator $B_P \equiv \prod_{l\in \partial P} \sigma_l^z$ measures the magnetic flux through the plaquette $P$.

The model is exactly solvable as both terms in $H_0$ commute with each other and the gauge constraint $G_s$. The ground state is charge-free and vortex-free: a simultaneous $+1$ eigenstate of $\tau_s^x$ and $B_P$ for all $s,P$. As the ground state is free of charge, it is a loop gas of electric flux. That is, it is an equal amplitude superposition of configurations where links with $\sigma_l^x=- 1$ form closed loops. The degeneracy of the ground state manifold depends on the topology of the lattice; on the torus, it is four-fold degenerate. The four ground states cannot be distinguished locally. They are labelled by the eigenvalues $\pm 1$ of the non-local Wilson loop operators $W^{nc} \equiv \prod_{l\in C_{nc}} \sigma_l^z$ along the two distinct non-contractible loops $C_{nc}$ on the direct lattice.

The elementary excitations of the model are gapped and are of two types: $e$ and $m$. $e$ denotes the presence of electric charge on site $s$, while $m$ is a magnetic vortex on plaquette $P$ characterized by $B_P= -1$. $e$ and $m$ are individually bosonic, but have mutual semionic statistics. The non-local string operators
\begin{align}
W^{e}(s, s') =\prod_{\substack{l \in C :  \\ {s,s'} \in \partial C}}  \tau_s^z \sigma_l^z\tau_{s'}^z \,,\quad W^{m}(\bar{s}, \bar{s}') =\prod_{\substack{l \in \bar{C} :  \\ {\bar{s},\bar{s}'} \in \partial \bar{C}}}  \sigma_l^x \label{Eq:Wm}
\end{align}
defined respectively on the curves $C$ and $\bar{C}$ on the direct and the dual lattice, create a pair of electric charges and vortices at their ends as shown in Fig.~\ref{Fig:TC}.

$H_{0}$ is robust to small local perturbations and extends to a topological phase. Nevertheless, a strong enough perturbation will eventually drive a transition into a trivial phase. The toric code Hamiltonian perturbed by transverse fields is $H=H_{TC} -  \sum_l \Gamma \sigma_l^x + J \sigma_l^z $; both perturbations drive continuous transitions to trivial (spin-polarized) phases when made large. In the gauge-invariant formulation of the \zt gauge theory with matter, the perturbed Hamiltonian takes the form 
\begin{align}
\label{Eq:IsingHam}
-H  &= K \sum_P B_P  + \Gamma_M \sum_s \tau_s^x + J\sum_{l} \sigma_l^z \prod_{s\in\partial l} \tau_s^z +  \Gamma \sum_{l} \sigma_{l}^x  .
\end{align}
The phase diagram of this theory was explained in detail in the seminal paper by Fradkin and Shenker \cite{FradkinShenker} and has more recently been confirmed in several numerical studies \cite{Dusuel, Tupitsyn, Vidal, Trebst, Wu:2012kx}. We will now briefly review this model in different parameter regimes.

\subsection{Pure matter theory ($\Gamma = 0$)}  \label{Sec:PureMatter}
For $\Gamma = 0$, the gauge degrees of freedom are static and frozen into a vortex-free configuration in the ground state sector. It is therefore convenient to diagonalize $H$ in the gauge-variant subspace where $\sigma_l^z=1$ for all $l$ and project the eigenstates to the gauge-invariant subspace afterwards. The Hamiltonian then maps to the (2+1)D transverse field Ising model (TFIM) for the matter spins:
\begin{align}
\label{Eq:purematter}
H_{\scriptscriptstyle TFIM}= - J\sum_{ \langle s s'\rangle} \tau_{s'}^z \tau_s^z  - \Gamma_M \sum_s \tau_s^x.
\end{align}
On tuning $J$, the TFIM undergoes a conventional `Higgs' phase transition from a paramagnetic phase to a symmetry-broken ferromagnetic phase. In a complementary view, the static electric excitations $e$ defined at the toric code point ($\Gamma=J=0$) acquire dynamics when $J\neq 0$ and eventually condense at a critical value of $J$. The transition is in the $3D$ Ising universality class and is detected by the local order parameter, $\langle \tau^z\rangle$ in the gauge-variant subspace. In the gauge-invariant subspace, $\tau_z$ maps on to a non-local string operator.

After projection, the state with $\{\sigma_l^z=1\}$ is the vortex-free configuration in the topological sector defined by the Wilson loop $W^{nc}=1$ for both non-contractible loops on the torus. This choice maps to a TFIM with periodic boundary conditions in both directions of the torus; the remaining three topologically inequivalent vortex-free configurations generate TFIMs with different boundary conditions (periodic-antiperiodic etc.). The four-fold degeneracy of the topological phase vanishes in the Higgs/ferromagnetic phase.

\subsection{Pure Gauge Theory ($J = 0$)} \label{Sec:PureGauge}
In this case, matter is static. The ground state is in the charge-free sector ($\tau_s^x = 1$), and the Hamiltonian for the gauge variables in this sector is:
\begin{align}
H_{\mathbb{Z}_2} = -K \sum_P B_P - \Gamma \sum_{l} \sigma_{l}^x.
\label{Eq:puregauge}
\end{align}
When $\Gamma/K$ is small, the gauge variables are weakly fluctuating and the elementary excitations are well-described as vortex pairs. At some critical $\Gamma/K$, the vortices condense and the gauge variables strongly fluctuate past this point. This transition cannot be diagnosed by a local order parameter. Instead, the vortex condensate phase is marked by the vortex pair creation operator, $\langle W^m \rangle \neq 0$ for vortices separated by long distances. In the conjugate electric field basis, flux loops become costly as $\Gamma$ is increased; hence the transition is from a topological loop gas phase at the toric code point to a phase in which flux loops become confined.

In a different language, the transition is understood as a deconfinement-confinement transition for the static electric charge\cite{FradkinShenker} and is diagnosed by the free energy cost of creating a pair of (infinitely separated) charges. The cost is finite in the deconfined phase, but infinite in the confined phase, and is equivalent to the change in behavior of the expectation of the contractible Wilson loop:
\begin{align} \label{Eq:Wilson} W(L) \equiv \left\langle \prod_{l\in C} \sigma_l^z \right\rangle \end{align}
from a perimeter law ($W(L) \sim \exp(- L)$) to an area-law ($W(L) \sim \exp(-L^2)$). $C$ is a contractible loop and $L$ is its perimeter.

In $d=2$, the \zt Ising gauge theory is self dual\cite{Wegner, Kogut}. Thus the pure gauge theory also maps to a (2+1)D TFIM and the confinement-deconfinement transition belongs to the 3D Ising universality class. The details of this duality are explained in Ref. \onlinecite{Kogut}, and have been summarized in Fig.~\ref{Fig:duality}. We emphasize that despite the duality, the transition is \emph{not} described by a local order parameter, as will be even clearer in Sec.~\ref{Sec:LW}.

\subsection{The full phase diagram}
\begin{figure}
\includegraphics[width=8.3cm]{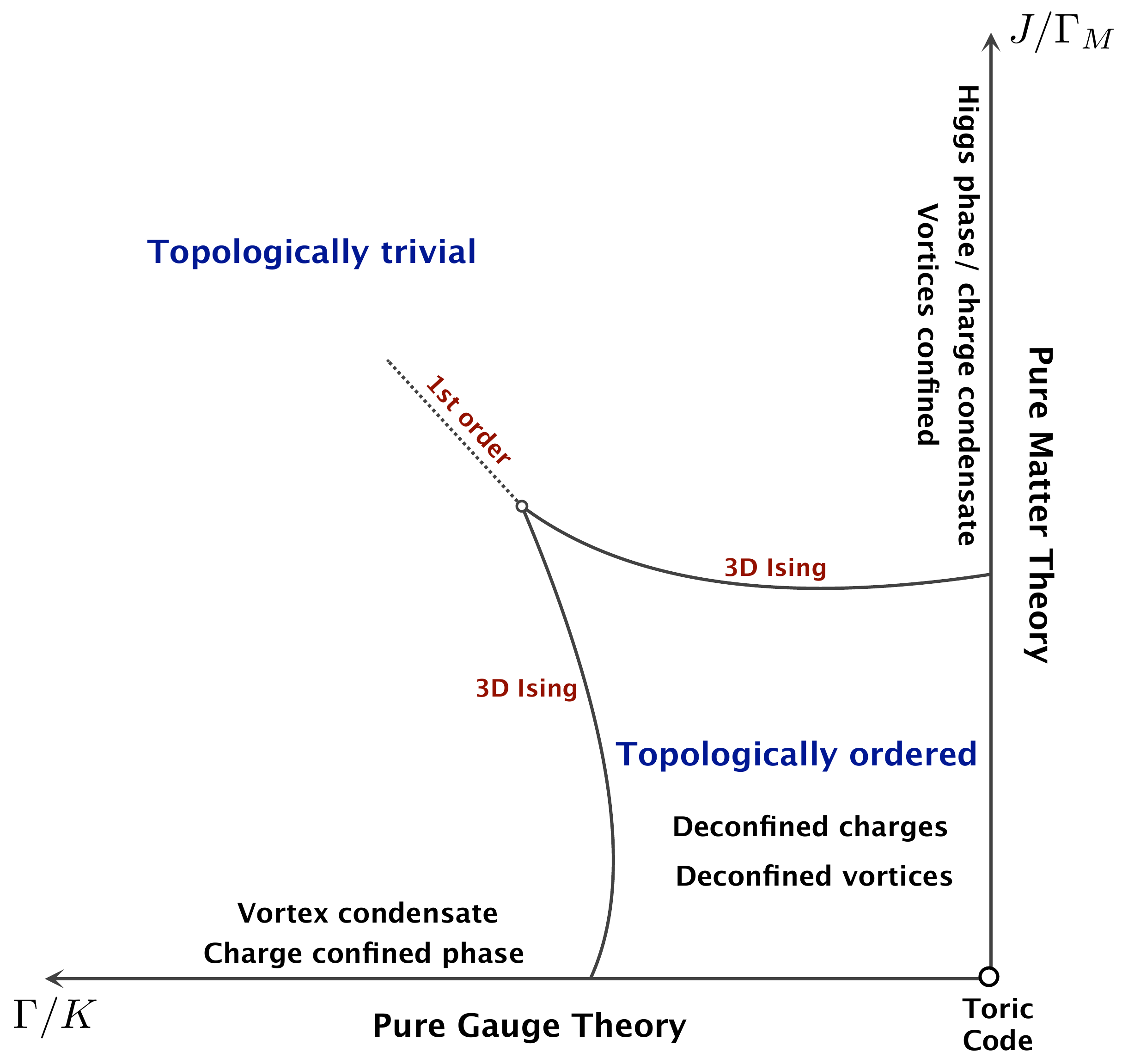}
\caption{$T=0$ phase diagram of the \zt theory in $d=2$ dimensions. The matter and gauge axes are dual, and the Higgs and charge confined phases are smoothly connected. }
\label{Fig:PhaseD}
\end{figure}

The full $T=0$ phase diagram of the Ising gauge theory in (2+1)D is shown in Fig.~\ref{Fig:PhaseD}. Fradkin and Shenker\cite{FradkinShenker} have shown that the confinement/Higgs transitions are stable on moving away from the pure gauge/matter axes. Further, the Higgs and confined phases are smoothly connected. However the diagnostics previously discussed, like the Wilson loop, no longer differentiate between the two phases. In a recent paper\cite{ShivajiDeconfinement}, Gregor et. al. have shown that an appropriately defined line tension, related to the Fredenhagen Marcu\cite{FM, FM2} order parameter studied by lattice gauge theorists, can be used to diagnose the transition everywhere in the phase diagram. We will use this quantity to study ramps across generic points on the critical line in the phase diagram.

\section{Kibble Zurek I - Ramp across the Higgs transition}
\label{Sec:KZ1}
We begin by reviewing the Kibble-Zurek (KZ) formalism for linear ramps (a Trans-Critical-Protocol in the parlance of Ref.~\onlinecite{Chandran}) across the conventional 3D Ising transition along the pure matter line at $\Gamma = 0$ in Sec.~\ref{Sec:KZ1Review}. The late time evolution of the system is naturally described as ``coarsening": a dynamical process previously discussed only in classical systems. We take the first steps to apply these ideas to an isolated quantum system in Sec.~\ref{Sec:KZ1Coarsening}.

\subsection{Scaling theory: Review}
\label{Sec:KZ1Review}
The system is in equilibrium in the paramagnetic phase at $t = -\infty$ and is driven to the ferromagnetic phase by changing the transverse field $\Gamma_M(t)$ linearly: 
$\delta(t) \equiv (\Gamma_M(t) - \Gamma_{Mc})/\Gamma_{Mc}  = -t/\tau$. The critical point (CP) is at $\delta = 0$ and $\tau$ is the ramp time. The response of the system to slow ramps is characterized by three chronological regimes: adiabatic evolution at early times, `critical' or diabatic evolution near the CP, and a late-time regime that we argue to be domain-growth.

At early times, the system is far from the CP and evolves adiabatically. Critical slowing down implies that the instantaneous correlation time diverges as $\xi_t \sim |\Gamma_M-\Gamma_{Mc}|^{-\nu z}$ near the CP and adiabaticity must break down before the CP is reached. $\nu$ and $z$ are respectively the correlation length and dynamic exponent with $\nu=0.627$ and $z=1$ in this case\cite{Ferrenberg:1991pi} . The system falls out of equilibrium at the KZ time, $t= -\tk$, when the time remaining to reach the critical point, $\tk$, becomes equal to $\xi_t$:
\begin{align}
\label{Eq: tKibble}
\xi_t(-\tk; \tau) = \tk  &&\Rightarrow\,\, \tk = \tau^{\frac{\nu z}{\nu z + 1}}.
\end{align}
At $t = - \tk$ the evolution of the system become diabatic and, to zeroth order, the system remains frozen until it emerges on the other side of the CP at $t = \tk$. The KZ time, $\tk$, defines a KZ length, $\lk = \tk^{1/z}$ which is the correlation length at the time the system falls out of equilibrium. Finally, for $t\gg \tk$, we get coarsening.

Recently, Chandran et. al.\cite{Chandran} formulated a scaling limit in which the physics described above becomes universal for a given pairing of a critical point and a ramp protocol. This limit is defined as $\tau \rightarrow \infty$ with time and length scales measured in units of the diverging scales $\tk$ and $\lk$. As $\delta(\tk) \rightarrow 0$ in this limit, the out-of-equilibrium response of the system is completely controlled by the critical point. The content of the scaling theory is not just the critical exponents, previously discussed by Kibble and Zurek, but also scaling functions for various physical observables. For example, the equal-time two-point correlation function of the order parameter defined as:
\begin{align}
\label{Eq:Gtt}
\langle \tau^z_{s}(t) \tau^z_{s'}(t)\rangle_{\tau} \equiv G_{\scriptscriptstyle \tau \tau}(|\mathbf{s}-\mathbf{s}'|, t;\tau),
\end{align}
has the KZ scaling form 
\begin{align}
\label{Eq:KZscaling}
 \lim_{\tau \rightarrow^* \infty} \lk^{2\Delta} G_{\scriptscriptstyle \tau \tau}(x, t) = \mathcal{G}_{\scriptscriptstyle \tau \tau}\left( \frac{x}{\lk}, \frac{t}{\tk} \right)
\end{align}
where $\tau \rightarrow^* \infty$ is defined as the limit $\tau \rightarrow \infty$ with $\frac{x}{\lk}$ and $\frac{t}{\tk}$ held fixed, $x$ is the distance between sites $s$ and $s'$, and $\Delta$ is the scaling dimension of the operator $\tau^z$. $\Delta$ is equivalent to $\beta/\nu$, and is numerically found to be $0.518$ in the (2+1)D TFIM\cite{Ferrenberg:1991pi} .

The scaling function should asymptote to the correct equilibrium form in the limit $t/\tk \rightarrow -\infty$ with $x/\xi(t;\tau)$ fixed 
\begin{align}
\label{Eq:equilibrium}
\mathcal{G}_{\scriptscriptstyle \tau\tau}\left( \hat{x}, \hat{t}\right) &\sim {\hat{t}}^{2\nu\Delta} \mathcal{G}^{eq}_{\scriptscriptstyle \tau \tau}\left(\hat{x}{\hat{t}}^{\nu}\right)
\end{align}
where $\hat{x}$ and $\hat{t}$ are defined as $x/\lk$ and $t/\tk$ respectively, $\mathcal{G}^{eq}$ is the equilibrium scaling function, known to decay exponentially in the Ising model, and $\hat{x}\hat{t}^\nu = x/\xi(t)$. \emph{For the rest of this article, the hat superscript will be reserved for the variables scaled by $\lk$ or $\tk$, depending on their units.}

A full description of the diabatic regime, previously termed `critical coarsening'\cite{Biroli}, is much harder. The difficulty surpasses static computations in the analogous quantum critical regime as it involves real time.

As the scaling content at late times has not appeared in the literature before, we devote the next subsection to it. We can also investigate the scaling functions for thermodynamic quantities like the excess energy density above the ground state \cite{De-Grandi:2010aa}, $q$, and the entropy density\cite{Chandran, Anatoli:2011aa} $s$:
\begin{align}
\label{Eq:heat}
q(t;\tau) &\sim \lk^{-d} \,\tk^{-1} \,\mathcal{Q} (\hat{t}) \\
s(t;\tau) & \sim \lk^{-d}\, \mathcal{S}(\hat{t})
\end{align}
Both quantities tend to zero as $t/\tk \rightarrow -\infty$ when the evolution is adiabatic and become non-zero when the system falls out equilibrium. The evolution at late times is therefore best understood through the finite excess energy density ($q>0$) or finite temperature phase diagram. 

\subsection{Coarsening}
 \label{Sec:KZ1Coarsening}
We now address the late time dynamics of the KZ ramp. It is generally believed that a \emph{classical} system quenched to an ordered phase with multiple vacua undergoes coarsening, whereby each local broken-symmetry region grows in time and the system is asymptotically statistically self-similar on a characteristic length scale, $l_{\rm co}(t)$. Put another way, the two-point function heals to its equilibrium value on the scale $\xi$ within each ÒdomainÓ, and is exponentially suppressed between domains, each of growing length $l_{\rm co} \gg \xi$. In the late time regime, dynamical scaling is expected to hold when there are no growing scales competing with $l_{\rm co}$. For more details, see Ref.~\onlinecite{Bray}.

We now generalize this idea to the KZ ramp in the \emph{quantum} TFIM. For simplicity, let us stop the ramp in the ordered phase at some $t/\tk = \hat{t}^s \gg 1$, while continuing to measure time and length on the scales set by $\tk, \lk$. The superscript $s$ denotes stopping. The system initially appears disordered. At infinitely long times however, we expect that the system is thermal and ordered, as the (2+1)D TFIM is not known to be integrable.  Further, we expect the approach to equilibrium to be through domain growth or coarsening, driven by the lack of long range order at late times. The system then locally breaks the symmetry but is globally disordered, with long domain walls past the growing length scale, $l_{co}^s(t)\gg \lk$. Assuming local equilibration, the physics of coarsening can be captured in a hydrodynamic theory with two slow modes: the non-conserved, scalar order parameter and the conserved energy density\footnote{The momentum density appears as an additional conserved quantity in this field theory. It is our current belief that this does not change the relevant power in the coarsening regime, but we are investigating this.}. We shall call this Model C in a slight abuse of language (properly it refers to the theory with thermal noise included \cite{Hohenberg}). As $\hat{t}\rightarrow \infty$ , we therefore predict that the system obeys the dynamic scaling hypothesis, that is, it looks self-similar on the scale of a growing length $l^{s}_{co}(t;\tau) $ and that $\mathcal{G}^{s}_{\scriptscriptstyle \tau\tau }$ has the late time form:
\begin{align}
\label{Eq:stopcoarsening}
\mathcal{G}^{s}_{\scriptscriptstyle \tau\tau }\left( \hat{x}, \hat{t}\right) &\sim (\hat{t}^s)^{2\nu\Delta} \mathcal{G}^{co}_{\scriptscriptstyle \tau\tau}\left(\hat{x}\lk/l^{s}_{co} \right), \\
\textrm{where } l^{s}_{co}(t;\tau) &= \lk \left(\frac{t}{\tk}\right)^{1/z_d}\,\, \textrm{and } z_d=2. \nonumber
\end{align}
The value of the dynamic exponent, $z_d$, quoted above is only known numerically\cite{Kockelkoren,Kockelkoren2}. $\mathcal{G}_{\scriptscriptstyle \tau\tau }^{co}$ is a scaling function that can also be computed within Model C  \cite{Kockelkoren,Kockelkoren2, Zheng, SanMiguel} .

The above discussion hinges on two key assumptions. First, the infinite time state of the system should have long-range order, that is, the late time evolution should be in the ordered phase. More precisely, we require the excess energy density \eqref{Eq:heat} at the stopping time $q(\hat{t}^s;\tau)$ to be smaller than critical energy density $q_c(\hat{t}^s; \tau)$, below which the system will be ordered in equilibrium. The dominant contribution to $q$ (at $\hat{t}^s$) is from the defect density on the scale $\lk$ frozen in at $t\approx \tk$. Assuming that these defects evolve adiabatically for $t> \tk$, we may conservatively estimate $q$ to scale as the single-particle gap $1/\xi(\hat{t}^s; \tau)^z$. As promised, this density is much smaller than the instantaneous critical density,  $q_c \sim 1/\xi^{d+z}$:
\begin{align}
\label{Eq:Tfromqlateform}
\frac{q}{q_c} \sim \xi^d \sim  \left (\frac{1}{\hat{t}^s}\right)^{\nu d} \ll 1.
\end{align} 
For all $\hat{t}> \hat{t}^s$, the energy of the system is conserved and $q$ and $q_c$ do not change in time. Thus, the system evolves in the ordered phase at late times. 

The second assumption is that of local equilibration over hydrodynamical time-scales. In Model C, the latter is the time-scale for domain growth by $\lk$. It can be inferred from Eq.~\eqref{Eq:stopcoarsening} to be $\delta t_{co} \sim \tk \hat{t}^{1-1/z_d}$. On the other hand, the time-scale for local equilibration processes on the scale $\lk$ is set by $\tk$. The validity of Model C as the late-time dynamical description relies on the equilibration time being much smaller than $\delta t_{co}$. As the inequality $\tk \ll \delta t_{co}$ is parametrically controlled by $\hat{t}$, the coarsening behavior in Model C is a better and better approximation to the quantum dynamics as $\hat{t}\rightarrow \infty$.

Finally, in the original KZ problem (where we don't stop the ramp), the late time evolution is also in the finite-temperature ordered phase as the relation $q/q_c \ll 1$ holds for every $t/\tk \gg 1$. However, the continuously changing parameter in the Hamiltonian affects the local equilibration argument in two important ways. First, the characteristic size of the domains grows at a slower rate:
\begin{align}
\label{Eq:lcogrow}
l_{co}(t;\tau) =\lk \left(\frac{t}{\tk}\right)^\theta\,\textrm{where } \theta= \nu \left(\frac{z}{z_d} - 1\right) + \frac{1}{z_d}.
\end{align}
In the (2+1)D TFIM, $\theta = (1-\nu)/2$ and is smaller than $1/z_d$. This slowing down can only help in the argument given above. The second effect is that the single particle gap $\Delta$ grows as $1/\tk (t/\tk)^{\nu z}$ at late times. This, however, increases various scattering times  (and consequently various equilibration times) in the problem and the applicability of hydrodynamics here becomes a delicate affair. In Appendix~\ref{App:ToCoarsenorNotto}, we argue that the process that drives coarsening and increases entropy involves the interaction of the long domain walls with the bulk quasiparticles within each domain. As the bulk quasi-particles scatter off the walls parametrically many times before the system parameters are changed, coarsening can at least self-consistently be justified. We therefore conjecture that the two point function as $\hat{t}\rightarrow\infty$ holding $x/l_{co}(t)$ fixed obeys dynamic scaling: 
\begin{align}
\label{Eq:coarsening}
\mathcal{G}_{\scriptscriptstyle \tau\tau }\left( \hat{x}, \hat{t}\right) &\sim {\hat{t}}^{2\nu\Delta} \mathcal{G}^{co}_{\scriptscriptstyle \tau\tau}\left(\hat{x}\lk/l_{co} \right).
\end{align}

In this process, the entropy density increases weakly in time. The late time asymptotes reflect this:
\begin{align}
\label{Eq:LateTimeqs}
\mathcal{Q}(\hat{t}) &\sim q_0\hat{t}^{\nu z} + q_1 \hat{t}^{\nu z - \theta} \\
\mathcal{S}(\hat{t}) &\sim s_0 - s_1 \hat{t}^{ - \theta}. \nonumber
\end{align}
The leading terms in $\mathcal{S}$ would be present even if the evolution were adiabatic. The sub-leading term is the thermodynamic signature of coarsening. From this point, every time we invoke results from coarsening, the reader should keep in mind the subtleties presented in this section. 

\section{Kibble Zurek II - Ramp across the confinement transition}
\label{Sec:KZ2}
We now ramp across the pure gauge theory Eq.~\eqref{Eq:puregauge} by tuning $\Gamma$. In this case, the transition is from a topologically ordered deconfined phase to a confined one, and there is no description in terms of a local order parameter. Nevertheless, we will now show that the KZ mechanism for Landau transitions discussed in the previous section can be generalized to these transitions. Additionally, the loops and strings characterizing the topological phase (string-nets) will coarsen.

\begin{figure}
\includegraphics[width= 8.3cm ]{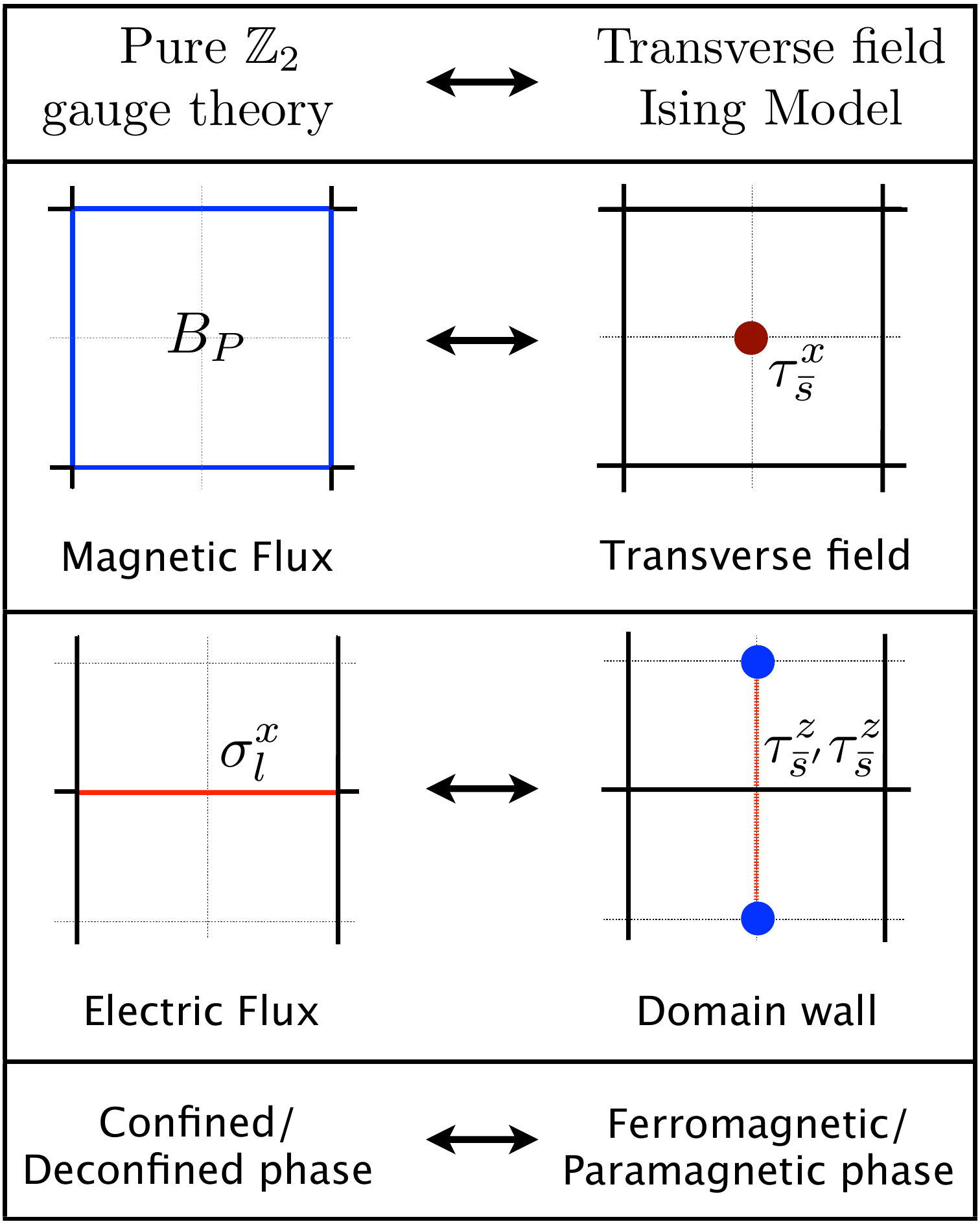}
\caption{Table summarizing the duality between the pure \zt gauge theory (\ref{Eq:puregauge}) and the TFIM in $d=2$. Dark and light lines denote the direct and dual lattice respectively.}
\label{Fig:duality}
\end{figure}

Our main tool is the duality in $(2+1)$D between the pure gauge theory and the TFIM summarized in Fig. \ref{Fig:duality}. Importantly for us, the presence of electric flux on a link (of the direct lattice) maps to a domain wall between the TFIM spins (on the dual lattice), while the vortex operator $B_P$ maps to the dual transverse field. The duality also ensures that a finite temperature confined phase exists, and that coarsening is described by the hydrodynamics of Model C.

For specificity, we begin the ramp at the deconfined toric code point in one of the ground state sectors. The ground state is a loop gas of the electric flux lines in the $\sigma^x$ basis. By duality, these are the domain walls of the paramagnetic phase of the TFIM. The system falls out of equilibrium in the deconfined phase before it is taken through the transition, with a network of loops of minimum size $\lk$. In the confined phase, flux loops map to the costly domain walls of the dual ferromagnetic phase. Post the diabatic regime in the confined phase, this network of loops (string-nets) is diluted (average size increases as  $l_{co}$) as the system coarsens. More generally, we can imagine string-nets being diluted in a generic topological theory and we will show some examples of this in Section \ref{Sec:LW}.

As before, energy conservation requires the decreasing electric flux density to be compensated for by an increasing bulk energy density. Essentially, the system arrives in the confined phase (which is a vortex condensate) with a greater electric field density and a smaller magnetic vortex density as compared to the instantaneous ground state. The subsequent evolution through coarsening increases the typical size of the electric flux loops to $l_{co}(t)$, thereby decreasing the electric field density and increasing the bulk energy density of the vortex condensate.

\begin{figure*}
\includegraphics[width = 17 cm]{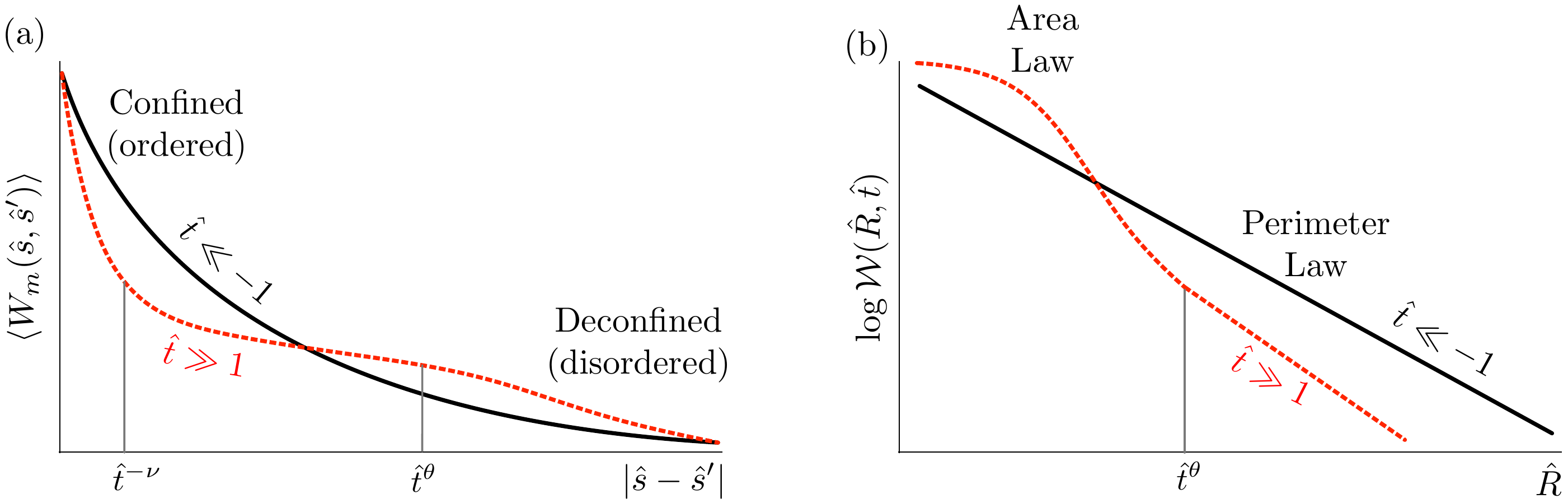}
\caption{ An illustration of two scaling functions showing adiabatic behavior in the deconfined phase at early scaled times (black)  and coarsening behavior in the confined phase at late scaled times (red). The crossover on the scale $\hat{l}_{co}\equiv l_{co}/\lk$ in the red curves is a signature of string-net coarsening. The hat superscript denotes scaled variables like $\hat{t}=t/\tk$ etc. (a) The scaling function of the string operator that creates a pair of vortices at $\hat{s}$ and $\hat{s'}$ ($\langle{W^m}\rangle $ defined in Eq.~\eqref{Eq:Wm}) as a function of the scaled vortex separation $|\hat{s}-\hat{s}'|$. This operator is dual to the two-point correlator $\langle \tau^z_s \tau^z_{\bar{s}'}\rangle$ in the TFIM. (b) The logarithm of the scaling function of the Wilson loop as a function of the scaled radius illustrating Eq. \eqref{Eq:WilsonScale}. The time dependence of $\hat{\xi}$ and $\hat{l}_{co}$ is respectively $\hat{t}^{-\nu}$ and $\hat{t}^\theta$.}
\label{Fig:vortexanndwilson}
\end{figure*}

Next, the two-point correlator that detects long-range order in the dual TFIM, $\langle \tau^z_{\bar{s}} \tau^z_{\bar{s}'} \rangle$,  maps to the vortex pair creation operator (\ref{Eq:Wm}), $\langle W^{m}(\bar{s}, \bar{s}')\rangle $ that detects vortex condensation. As the condensed phase is also a confining phase for charge, a non-zero value of $\langle W^{m}\rangle$ for long strings detects charge confinement. The scaling form for $\langle W^m\rangle$ is given by Eq.~\eqref{Eq:KZscaling}, and its asymptotic behavior is identical to that of the two-point function discussed in the pure-matter theory. In particular, in the coarsening regime, the dual TFIM is ordered on length scales less than $l_{co}(t)$. Correspondingly, $\langle W^{m} \rangle$ is also non-zero on scales shorter than $l_{co}(t)$ but decays exponentially on longer length scales. Thus, the non-local string operator $\langle W^{m} \rangle $ probes the crossover scale from confinement to deconfinement as a function of time. Fig. \ref{Fig:vortexanndwilson}(a) shows the scaling for $\langle W^{m}\rangle $.

Finally, we can consider an interesting observable that we did not discuss in the TFIM. This is the Wilson loop~(\ref{Eq:Wilson}), $W(R, t; \tau)$ on a curve of radius $R$. Were the evolution to be adiabatic, $W(R, t; \tau)$ for large $R$ would obey a perimeter law when $t<0$ and an area law when $t>0$. In the KZ scaling limit, the scaling of the Wilson loop takes the form $W(R, t; \tau) \sim \mathcal{W}(\hat{R} ,\hat{t})$, where $R/\lk = \hat{R}$. Its asymptotic behavior is:
\begin{equation}
\label{Eq:WilsonScale}
\mathcal{W}(\hat{R}, \hat{t}) \sim
\begin{cases}
\exp(-\hat{R} \,\hat{t}^{\nu}), & \mbox{if } \hat{t} \ll -1 \\
\exp(-(\hat{R} \,\hat{t}^{\nu})^2), &\mbox{if } \hat{t}\gg 1 \mbox{ and }  \hat{R} \ll \hat{t}^\theta \\
\exp(-(\hat{R}/ \hat{t}^{\theta}) ), &\mbox{if } \hat{t}\gg 1 \mbox{ and } \hat{R} \gg \hat{t}^\theta.
\end{cases}
\end{equation}
These scaling forms follow simply from the picture of adiabatic evolution when $\hat{t} \ll -1$ and a growing length $l_{co}(t)$ separating confinement from deconfinement when $\hat{t}\gg 1$. The Wilson loop therefore also probes the crossover scale from confinement to deconfinement as a function of time. Fig.~\ref{Fig:vortexanndwilson}(b) shows the scaling of the Wilson loop.

\section{Kibble Zurek III - Ramp across a generic transition in the \zt theory}
\label{Sec:KZ3}
We will now see how the discussions of the previous two sections can be generalized to ramps crossing any critical point in the full \zt phase diagram. First, consider moving off the pure gauge line by introducing a small, but non-zero $J$. The coupling to matter is irrelevant to the $T = 0 $ transition; hence, the confinement-deconfinement transition in Fig.~\ref{Fig:PhaseD} persists for non-zero $J$ and remains in the same universality class. Since the gap to charge excitations does not close on making $J$ non-zero, we can re-write the Hamiltonian as one with no gauge-matter coupling (to any fixed order in $J$) through a canonical unitary transformation. The transformation defines ``dressed'' charge and gauge operators $-$ in the dressed variables, the ground state is charge-free and $\Delta_c$ is the non-zero gap to charge excitations.

As we heat the system in the process of the ramp, we also need to consider the finite temperature phase diagram and the excited spectrum when $J\neq 0$. Although the ground state is (dressed) charge-free, the excited states have an exponentially small density of charge, $e^{-\Delta_c/T_{eff}}$, at any effective temperature $T_{eff}$ corresponding to an excess energy density $q$. The presence of charge at finite temperatures is extremely significant for the late-time coarsening picture for two reasons. First, it destroys the finite temperature confined phase at any non-zero $J$, without which a coarsening description is not meaningful. Synergistically, a finite density of charge implies that the electric field lines naturally end somewhere. Thus, the pictures of flux-loops/domain walls coarsening are no longer sensible at the longest length scales.

Fortunately, the Kibble-Zurek scaling limit saves us from the problems raised above. This is because the ratio $q/q_c$ goes to zero as $t/\tk \rightarrow \infty$ (Eq.~\eqref{Eq:Tfromqlateform}) or equivalently, the effective local temperature, $T_{eff}$, computed from $q$ goes to zero in the scaling limit ($T_{eff}$ is well-defined as the system is locally in equilibrium. See Sec.~\ref{Sec:KZ1Coarsening}).  $\Delta_c$, on the other hand, remains finite. This implies that the ratio of the average distance between charges to the KZ length, $e^{\Delta_c/T_{eff}}/\lk$, is formally infinite in the scaling limit. Thus, while the dressed charges modify the \emph{true} long-time behavior by ending coarsening, the scales on which they do so lie outside the KZ scaling regime: in this way, the coupling to matter is a dangerously irrelevant variable in the KZ problem (in the scaling limit).

While we can write scaling functions for dressed observables, the results are not very elegant since the Hamiltonian dependent dressed operators are different at different points in time. A crisper solution is to use the line tension\cite{ShivajiDeconfinement}/ Fredenhagen-Marcu (FM) order parameter\cite{FM, FM2} alluded to previously. This is defined as
\begin{align}
\label{Eq:FM}
R(L) = \frac{W_{1/2}(L)}{\sqrt{W(L)}} = \frac{\langle \tau_s^z(\prod_{l\in C_{1/2}}\sigma_l^z)\tau_{s'}^z\rangle}{\sqrt{\langle \prod_{l\in C}\sigma_l^z\rangle}},
\end{align}
where $C$ is a square loop of side $L$ and $C_{1/2}$ is the open rectangle of sides $L$ and $L/2$ obtained by cutting $C$ in half, $s, s'$ are the endpoints of $C_{1/2}$ and $W$ is the contractible Wilson loop.

As $L\rightarrow\infty$, $R(L)$ is zero in the deconfined phase and non-zero otherwise. In this way, $R(L)$ acts as a test of long range ``order'', and appropriately generalizes the two-point spin correlator $G_{\tau\tau}$ from the pure matter theory (\ref{Eq:Gtt}), and the vortex pair creation operator $W^{m}$  \eqref{Eq:Wm} from the pure gauge theory. In fact, in the gauge-variant subspace $\{\sigma_l^z =1\}$ on the pure matter line, $R(L)$ exactly reduces to $G_{\tau\tau}$. The scaling form and asymptotes of $R(L)$ therefore follows from Eq.~\eqref{Eq:KZscaling} and the discussion below it. Of course, by duality, an identical analysis can be carried out by perturbing away from the pure-matter line as long as we interchange the gauge and matter degrees of freedom.

\section{Extension to generalized Levin-Wen models}
\label{Sec:LW}
In this section, we generalize the Kibble-Zurek problem to transitions in which topological order is reduced as opposed to destroyed. Specifically, we consider transitions out of a broader, non-Abelian, class of topological phases in lattice spin models of the Levin-Wen\cite{LW} type. Along special lines in the phase diagram, we show that the dynamics and scaling properties are exactly equivalent to those of the $\mathbb{Z}_2$ gauge theory. We identify analogous observables and the coarsening degrees of freedom of the string net that is condensed in the starting topological phase. However, we will see that the mapping of the dynamics is not an equivalence. We then consider perturbations away from this line, finding that as for the $\mathbb{Z}_2$ gauge theory, in the scaling limit these other perturbations do not alter the coarsening dynamics, but can be either irrelevant or dangerously irrelevant perturbations.

We restrict our discussion to the subset
of SU(2)$_k$ models whose topological order is that of a doubled, achiral,  Chern-Simons theory with gauge group SU(2) and a coupling constant of $k$ in appropriate units, though the construction of Ref. \onlinecite{LW} is more general. We also restrict to particular transitions that change the topological order by condensing bosonic vortex defects; the transitions we consider here were shown\cite{Burnell:2011kx} to be dual (in a certain limit) to the TFIM.

\subsection{Levin-Wen Hamiltonians with Ising transitions}

The SU(2)$_k$ models we study live in a Hilbert space built from tensoring a finite set of spin variables on each link of a honeycomb lattice, $\sigma_l~\in~\{ 0,\frac{1}{2}, 1,..., \frac{k}{2} \}$. These are analogous to the set of possible electric fluxes ($\sigma_l^x = \pm 1$) in the \zt gauge theory.
The idealized Levin-Wen Hamiltonians are similar in spirit to the toric code, and are constructed from a set of commuting projectors
\begin{align}
\label{Eq:HLW}
H_{\scriptscriptstyle LW} = - K \sum_P \mathcal{P}_P - \Gamma_M \sum_s \mathcal{P}_s
\end{align}
where $P$ represents a plaquette, and $s$ a site. These models can be viewed as deformations of lattice gauge theories with a continuous gauge group. They are `deformed' in the sense that their representation theory is truncated, even though the gauge group is not discrete. In our context, a lattice $SU(2)$ theory would have electric fluxes corresponding to all allowed spin values $0, 1/2, 1, ...$, while the models in question have a maximum spin $k/2$. (We refer to the link spins as `electric flux' though, more accurately, they are the representations of the lattice gauge/quantum group).
Instead of describing our analysis for general values of $k$, we will now specialize to $k=2$ in the interests of pedagogical simplicity and return
to comment on the generalization subsequently. 

We now discuss the detailed form of the Hamiltonian (\ref{Eq:HLW}) for SU(2)$_2$.
The vertex projector $\mathcal{P}_s$ penalizes violations of angular momentum conservation, analogous to the Gauss's law constraint $G_s =1$ in the \zt theory. If the three links entering a vertex have spins $i, j$ and $l$, angular momentum conservation requires $l \in i \times j$.  The rules for adding angular momentum have to be modified to be consistent with the truncation, however.  
For the SU(2)$_2$ model, the result is\cite{Bonderson}:
\be \label{Eq:AsU2}
\mc{P}_s | \includegraphics[height=.35in]{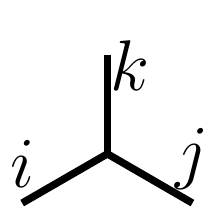} \rangle = \begin{cases} 1 \ \ \ &
\includegraphics[height=.35in]{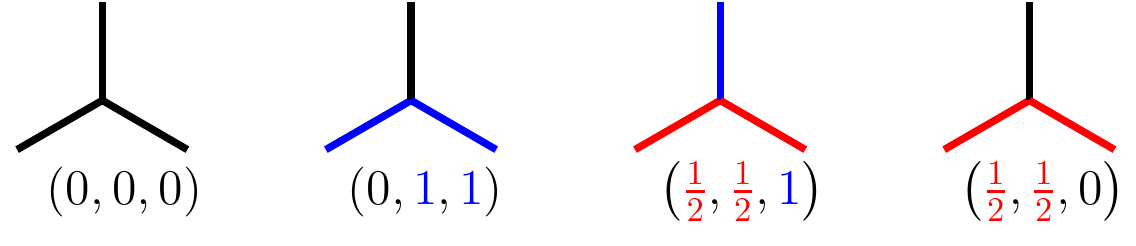} \\
0 & \text{otherwise}
\end{cases}
\ee
where it is understood that the eigenvalue of $\mc{P}_s$ is independent of interchanging the spins on the three links entering the vertex.

The plaquette term $\mc{P}_P$ projects onto states in which $P$ has no magnetic flux, and is written as a superposition of ``raising operators'': $\mc{P}_P = \frac{1}{\mc{D}} ( \mathbf{1}  + \sum_{\sigma = 1/2}^{k/2} a_\sigma B^\sigma_P)$, where $B^\sigma_P$ raises all spins on the plaquette $P$ by $\sigma$ in the truncated spin space. By ``raising'' , we mean a combination of raising and lowering angular momenta in the truncated spin space. $\mc{D}$ is the total quantum dimension, equal to $2$ here, while the coefficients $a_\sigma$ depend on the quantum dimension\cite{LW} of the spin representation $\sigma$. In $SU(2)_2$, they are $a_0 = 1, \; a_{1/2} = -\sqrt{2}, \; a_{1} = 1$. $B_P^\sigma$ raises all spins in P by raising the spin on each link $l \in \partial P$. The action of $B_l^\sigma$ on a link $l $ with spin $i \in \{0,1/2,1\}$ is:
\begin{align*}
 B^i_l|0 \rangle  & = |i\rangle \\
B^{1/2}_l |1/2\rangle &\propto |0\rangle \pm |1\rangle \qquad B_l^{1/2}|1\rangle \propto |1/2\rangle \\
B^1_l |1/2\rangle &\propto |1/2\rangle \qquad \qquad B^1_l |1\rangle \propto |0\rangle  
\end{align*}
The numerical coefficients are chosen such that the amplitude for creating any configuration with a $0$-eigenvalue under $\mc{P}_s$ is $0$, ensuring that the vertex and plaquette projectors commute. Their precise value is related to the $6j$ symbols of the quantum group $SU(2)_2$, but we will not require their detailed form here. Interested readers can consult Ref. \onlinecite{LW} for more details.

 As $\mc{P}_P$ and $\mc{P}_s$ commute, the spectrum of the Hamiltonian can be determined exactly. The ground state is a generalization (a string-net) of the loop gas ground state of the toric code, though there can be relative sign differences between terms in the Levin-Wen ground state wavefunction.
As in the toric code, the excited eigenstates of (\ref{Eq:HLW}) consist of ``matter" excitations of energy $\Gamma_M$, and ``vortex" excitations, of energy $K$.  In the SU(2)$_2$ model there are anyonic spin-$1/2$ charges,  fermionic spin-$1$ charges, and spin-$1/2$ or spin-$1$ vortices, both of which have bosonic statistics.  The spectrum can be made to correspond exactly to that of the doubled SU(2)$_2$ Chern-Simons theory.  Accordingly the topological ground state degeneracy is known\cite{Wu} to be $9$, as in the doubled Chern-Simons theory.

We can drive a phase transition in our system by perturbing the model 
(\ref{Eq:HLW}) with transverse fields which create pairs of charges or 
vortices, as we did for the toric code by adding $\sigma_l^x$ and 
$\sigma_l^z$.
The vortex excitations have bosonic statistics and hence transverse fields which create vortex pairs can drive a transition to a vortex condensed phase in which string-nets are confined. On the other hand, the analogue of the Higgs transition is not evident for our problem as both charges
are non-bosonic.

To drive the Ising transition that we are interested in, we add a transverse field which will condense spin-$1$ vortices.
The Hamiltonian that we will tune through this transition is
\begin{align}
 \label{Eq:SU22Trans}
H_{\text{SU(2)}_2}& =& -K  \sum_P  \frac{1}{2}  (\mathbf{1} + B^1_P)  - \Gamma_M \sum_s \mc{P}_s \n
&& -K \sum_P \; \frac{1}{\sqrt{2 } } {B}^{1/2}_ P -  \Gamma \sum_l (-1)^{2 \sigma_l }
\end{align}
where we have separated $\mc{P}_P$ into operators that ``raise'' spins by integer and half-integer amounts, and added a transverse field perturbation, $\Gamma (-1)^{2 \sigma_l }$. The transverse field creates a pair of spin-1 vortices on the plaquettes adjacent to $l$, and has eigenvalue $1$ on integer spin links, and $-1$ on half-integer spin links. Because the transverse field term squares to the identity (and all vortex creation operators commute), the vortices are Ising like.

On every plaquette $P$ and site $s$, the eigenvalues of $B_P^1$ and $\mc{P}_s$  are conserved, since these operators commute with $H_{\text{SU(2)}_2}$.
Thus, we can consider the transition engendered by varying the ratio $K / \Gamma$ in the subspace of the Hilbert space where the conditions
\begin{align}
\label{Eq:SU2restrict}
\mc{P}_s |\Psi \rangle = |\Psi \rangle, \qquad  B^1_P |\Psi \rangle = |\Psi \rangle
\end{align}
are always satisfied. In this subspace, the transition can be mapped onto the transition in the pure \zt gauge theory discussed in Sect. \ref{Sec:PureGauge}, and is therefore in the 3D Ising universality class as discussed in Ref.~\onlinecite{Burnell:2011kx}.
 Here we give a different derivation of this result which focuses on the ground state wavefunctions and is better adapted to our purposes in this
paper.

To understand the mapping between $H_{\text{SU(2)}_2}$ and  $H_{\mathbb{Z}_2}$ (\ref{Eq:puregauge}), notice first that the condition $\mc{P}_s|\psi\rangle = |\psi\rangle$ ensures that we are always working in the ``charge-free'' sector where the (deformed) angular momentum is conserved at each vertex. This,
together with Eq.~(\ref{Eq:AsU2}), stipulates that we only need to consider configurations where the number of half-integer spins entering each vertex is even - or equivalently, configurations in which half-integer spins form closed loops. Similarly, in the absence of charge in the pure \zt theory (\ref{Eq:puregauge}), the gauge constraint $G_s$ in Eq. \eqref{Eq:Gaugeinvariant} ensures that links with electric flux ($\sigma^x = -1$) form closed loops. The Levin-Wen transverse-field operator $(-1)^{2 \sigma_l}$ assigns an energy penalty to the spin $1/2$ edges that form these loops, similar to the action of the transverse-field term $\sigma_l^x$ on links with electric flux in the \zt theory.
% it easy to see that its anti-commutation relation with $B^{1/2}_P$ is identical to that of the plaquette and transverse field terms of Eq. (\ref{Eq:puregauge}).
{\it Thus both models describe a transition in which loops (of half-integer spin variables in the Levin-Wen case or $\sigma^x=-1$ variables in the \zt gauge theory case) become confined, and vortices become condensed as $\Gamma/K$ increases.}

There is, however, a qualitative difference between the operators $ \prod_{l \in \partial P} \sigma^z_l $ and ${B}^{1/2}_{P}$, both of which change the number of loops in a given configuration.  While $\sigma^z_l$ simply flips the spin on the link $l$, the operator ${B}^{1/2}_{P}$ maps a spin $0$ or $1$ link to a spin $1/2$ link, but a spin $1/2$ link to a superposition of a link in the state $0$ and a link in the state $1$.  Thus one might worry that the two operators generate the same set of configurations (after identifying $s = 0,1$ with $\sigma^x =1$, and $s= 1/2$ with $\sigma_x = -1$), but with different statistical weights.

We show in Appendix \ref{App:ProbApp} that this is in fact not the case.  Specifically, we prove that for any $\Gamma$, the ground state wave-function of either model can be expressed in the form
\be
|\Psi \rangle = \sum_{ \{ l \} } \beta^\Gamma_{\{ l \} }  | \Psi_{ \{l \} }  \rangle
\ee
where $ \{l \}$ denotes a set of links on which $\sigma_l = 1/2$ in the Levin-Wen model restricted to \eqref{Eq:SU2restrict}, or $\sigma^x_l =-1$ in the pure \zt gauge theory.   Crucially, we find that $ \beta_{\{ l \} }$ is the same for each set $\{l \}$ in both models.  Operators in the Levin-Wen model which commute with the conditions (\ref{Eq:SU2restrict}) are either diagonal in the vortex basis, or diagonal in the spin basis and sensitive only to the spin on each edge modulo $1$.   The expectation value of any such operator is therefore identical to that of its \zt analogue, cementing the equivalence of the two models.

We conclude that within the sub-sector \eqref{Eq:SU2restrict}, the transition is equivalent to that of the pure $\mathbb{Z}_2$ gauge theory, and dual to that of the TFIM.  It follows that our previous discussion of string net coarsening, and the scaling of vortex creation operators, applies {\it mutatis mutandis} to 
the model at hand.

Note, however, that the topological order of the initial and final phases of Eq. (\ref{Eq:SU22Trans}) is not the same as in the $\mathbb{Z}_2$ gauge theory; there are additional deconfined excitations on both sides of the transition.  (In fact, the confined phase of $H_{\text{SU(2)}_2}$ is a $\mathbb{Z}_2$ gauge theory\cite{TSBShort}).  There must therefore be some operators in the Levin-Wen model whose behavior through the ramp is not captured by the mapping to the Ising gauge theory.

To make this more explicit, we consider the fate of Wilson loop operators.  In the Levin-Wen model there are two of these:
\begin{align} \label{Eq:LWWL}
W_{1/2} (R,t; \tau) = \left\langle \prod_{l \in C} {B}^{1/2}_{l} (t)\right \rangle    \n
W_{1} (R,t; \tau) = \left\langle \prod_{l \in C} {B}^{1}_{l} (t) \right\rangle.
\end{align}
In order not to create vortices along the curve $C$, these raising operators must act with appropriate configuration-dependent complex coefficients, as discussed in Refs. \onlinecite{LW,CH}. $W_{1/2} (R,t; \tau)$ is clearly the analogue of the Wilson loop operator in the $\mathbb{Z}_2$ gauge theory; its expectation value obeys a perimeter law in the small $\Gamma/K$ phase, and an area law in the large $\Gamma/K$ phase, and its universal scaling in a linear ramp is given by Eq. (\ref{Eq:WilsonScale}).  However, as the spin-$1$ variable remains deconfined throughout the phase diagram, $W_{1} (R,t; \tau)$ always obeys a perimeter law.  Its expectation remains constant in the scaling limit as the system passes through the critical point.

Though we have primarily discussed the SU(2)$_2$ Levin-Wen model, the main results apply to a large family of models in which there is an excitation that behaves like the Ising vortex\cite{Burnell:2011kx}. Specifically, all the SU(2)$_k$ models exhibit Ising transitions in which the half-integer spins (integer spins) can be mapped onto \zt gauge configurations with $\sigma^x = -1$ ($\sigma^x = 1$).  They have two families of Wilson-line operators: the half-integral Wilson line operators, which obey an area law in the confined phase, and scaling relations analogous to those of Eq.~(\ref{Eq:WilsonScale}); and the integral Wilson line operators, whose expectation values do not depend on $t, \tau$ and which remain perimeter law throughout the ramp \footnote{The Higgs transition may survive when $k$ is a multiple of $4$ as some of the charges are bosons. Little is known about these transitions. }.

It is worth mentioning that the Ising transition we have discussed here is but one of a variety of confining transitions that can be realized in Levin-Wen models\cite{Burnell:2011kx,Gilsetal,VidalUnpublished}.  In the SU(2)$_2$ model discussed above, for example,  we could also add a transverse field term of the form $\cos \pi s_l$ (which has eigenvalues $(1, 0, -1)$ for $s_l = (0, 1/2,1)$, respectively).  This confines both spin-$1/2$ and spin $1$ labels, engendering a transition to a completely confined phase where both Wilson loop operators in Eq. (\ref{Eq:LWWL}) obey an area law. In this case the vortices that proliferate are not Ising-like, however, since the operator $\cos \pi s_l$ does not square to $1$.  Very little is known about the critical theory in this case, and we expect that the transition is not in the 3D Ising universality class, so that the scaling functions and coarsening behavior will be fundamentally different from those of the $\mathbb{Z}_2$ gauge theory.

\subsection{Away from the pure $\mathbb{Z}_2$ limit}
Thus far, we did not concern ourselves with the other excitations in the $SU(2)_2$ Levin-Wen model as their number was conserved in the ramp and we remained in the subspace (25) at all times. However, the other excitations, charges of spin 1 and 1/2, and vortices of spin 1/2, will be created in a ramp if we perturb away from the limit of Eq. (\ref{Eq:SU22Trans}) by adding terms to the Hamiltonian which break the local conservation laws. In particular, the spin 1/2 charge has anyonic statistics relative to the condensing vortices, so that a finite density of these destroys confinement (these are analogous to the matter sources of the Ising gauge theory). Once again, the KZ scaling limit saves the day. These spin 1/2 charges remain gapped throughout the transition. Thus, as for the Ising gauge theory with matter, in the scaling limit we expect that the density of all of these excitations is vanishingly small throughout the coarsening regime; terms violating the conservation of the spin-$1/2$ charge at each vertex then act as dangerously irrelevant variables in the manner described in Section \ref{Sec:KZ3}.

Spin $1$ charges and spin-$1/2$ vortices, however, have bosonic statistics relative to the spin $1$ vortex, and do not have analogues in the \zt theory.  Once again, for small perturbations which violate the exact local conservation of these excitations,  they remain gapped throughout the transition and hence do not affect coarsening in the KZ scaling limit.  Since a dilute density of such charges does not destroy confinement, however, we expect that they will not destroy coarsening even outside of the scaling limit.

\section{Concluding Remarks}
In this paper, we have initiated the study of the Kibble-Zurek problem for topologically ordered phases by studying the linear ramp across a transition that reduces/breaks topological order and written down a scaling theory for it. Interestingly, unlike broken symmetry cases where it is natural to ramp from less to more order, here it is more natural to ramp from more to less order. The latter leads to our identification of the slow dynamics of string net coarsening much as the former leads to defect coarsening {\it \`{a} la} Kibble and Zurek.  Of course, one {\it can} study the reverse protocol and the associated scaling although we have not done so here in the interests of not taxing the reader's patience unduly.

The basic framework here can be easily generalized to other transitions out of topological phases; although for string-net coarsening to be visible, the gauge degrees of freedom must have a ready identification. Examples are transitions out of $\mathbb{Z}_n$ phases with $n \ge 3$ in $d=2+1$ and with $n \ge 2$ in $d=3+1$. The Levin-Wen models also offer a ``target rich" domain, although the analysis is likely to prove more complicated for more general condensation transitions. It will also be interesting to move to contexts with conserved currents where one can study the temporal and spatial evolution of transport coefficients, such as the Hall conductance.

Finally, for the statistical mechanically inclined, we would like to draw attention to our identification of gapped matter as a dangerously irrelevant variable in the dynamical KZ context. This is clearly a more general idea---e.g. irrelevant departures from integrability will be similarly dangerous---and it suggests that in the KZ problem, more couplings will be classified as such than in the standard equilibrium analysis.

\section{Acknowledgements}
We are very grateful to A. Polkovnikov for his valuable comments on a draft of this article and specifically, for alerting us to the subtlety of the late-time coarsening process. We would also like to thank D. Huse for enlightening discussions on related issues. We are very grateful to A. Erez and S. S. Gubser for a stimulating collaboration on closely related work and to C.R. Laumann for many long and productive conversations. We would also like to acknowledge useful discussions with A. Rahmani, S. Vishveshwara and M. Kolodrubetz. A.C. and S.L.S. would like to thank KITP for generous hospitality during the finishing stages of this work. This research was supported in part by the National Science Foundation under Grant No. NSF PHY11-25915 and DMR 10-06608. 

\begin{appendix}

\section{Scattering times in coarsening}
\label{App:ToCoarsenorNotto}
Here, we identify the dynamical process enabling coarsening at late times alluded to in Sec.~\ref{Sec:KZ1Coarsening}, and justify that it remains in equilibrium during the KZ ramp. The criterion to remain in equilibrium is that the time-scale for such a process, $t_{co}$, is parametrically smaller than the time-scale for the change in the transverse-field $t_{\Gamma}$. Using $(\Gamma_M-\Gamma_{Mc})=-t/\tau$, we estimate $t_{\Gamma}$ to be:
\begin{align*}
t_{\Gamma} \equiv \frac{\Gamma_M-\Gamma_{Mc}}{d\Gamma_{M}/dt} = t.
\end{align*}
The system at late times has two kinds of excitations that are remnants of the paramagnetism at early times: 1) The long domain walls of average size  $l_{co}$  and 2) The bulk gapped quasi-particle excitations about each ferromagnetically ordered state. As $t/\tk \rightarrow \infty$, the latter can be treated as classical particles. The average density of these particles and their momentum is essentially determined at $t\sim \tk$ and is fixed to be $\sim 1/\lk^2$ and $1/\lk$ respectively. Their mass is determined by the gap $\Delta(t)$. The growing mass and the long inter-particle distances as compared to the instantaneous correlation length $\xi$ justify the classical particle approximation. An average velocity of these particles can be determined as
\begin{align*}
v_p \sim \frac{p}{m} \sim \frac{1/\lk}{\Delta},
\end{align*}
where $p$ is the average momentum and $m$ the mass. The mechanism of coarsening proceeds through the transfer of energy between the long domain walls and these particles. To wit, the relevant time-scale $t_{co}$ is the scattering time between these particles and the wall:
\begin{align*}
t_{co} \equiv \frac{l_{co}}{v_p}.
\end{align*}
Recall that the growth law when $\xi$ is a function of time is (Eq.~\eqref{Eq:lcogrow}):
\begin{align*}
l_{co}(t;\tau) \sim \xi(t;\tau)\left( \frac{t}{\xi(t;\tau)^z}\right)^{1/z_d} \sim \lk \left(\frac{t}{\tk}\right)^{\frac{1-\nu}{2}},
\end{align*}
where in the last step, we have substituted the critical exponents of the (2+1)D TFIM, $z=1$, $z_d=2$. Putting the pieces together, we see that $t_{co} \ll t_{\Gamma} \Rightarrow \nu < 1$. This certainly holds at the 3D Ising critical point where $\nu \approx 0.6$. Thus, we conclude that coarsening described by Model C is indeed the correct long time asymptote for the KZ scaling functions in a linear ramp.

Finally, we observe that all dynamical processes in the (2+1)D TFIM do not remain in equilibrium in the KZ ramp at late times. The scattering time between quasi-particles, $t_{pp}$, grows as $\Delta^2$ in this limit and is parametrically larger than $t_\Gamma$. A hydrodynamical description, if it exists, is therefore more delicate than the case when the ramp is stopped at some $t/t_K = \hat{t}^s$.   

\section{Mapping of general SU(2)$_k$ models to the $\mathbb{Z}_2$ gauge theory}
\label{App:ProbApp}
In this Appendix, we will discuss in more detail the mapping from the confining transition in the $SU(2)_2$ Levin-Wen model to the pure $\mathbb{Z}_2$ gauge theory (\ref{Eq:puregauge}). As discussed earlier, the transition in question involves varying $K/\Gamma$ in Eq.~(\ref{Eq:SU22Trans}), while restricting the Hilbert space to states with eigenvalue 1 under the vertex projector, and the integer part of the plaquette projector. In this subspace, links with half-integer spins form closed loops. The mapping to the pure gauge \zt theory involves mapping half-integer (integer) spins to the presence (absence) of \zt electric flux $\sigma^x = -1\; (+1)$.  The transverse field operator $(-1)^{2\sigma_l}$  maps to $\sigma^x_l $, and $B^{1/2}_P$ to $B_P = \prod_{l\in \partial P}\sigma_l^z$.   We will now show that the probability to be in any loop configuration is the same in both theories for every choice of $K/\Gamma$.

We will begin with some notation.  Let $C$ denote the collection of links on the lattice that form closed loops, and $\langle \alpha_{1/2}(C) \rangle $ the probability for a configuration in $C$.  We will work here in the restricted Hilbert space of states for which
\be \label{WhatIWant}
\mc{P}_s |\Psi \rangle = |\Psi \rangle \ \ \ \ \ \ \ \ \ \ \ B_P^{1} |\Psi \rangle = |\Psi \rangle
\ee
and assume that our lattice has no boundary. To make the analogy to the $\mathbb{Z}_2$ gauge theory, we also define the analogous operator, $\alpha_{x}(C)$, whose expectation value gives the probability for a closed loop configuration of links with $\sigma^x_l =-1$.

Our objective is to prove that, for every $\Gamma$ and $K$,  $\langle \alpha_{1/2}(C) \rangle = \langle \alpha_{x}(C) \rangle$.   Since operators that commute with the conditions (\ref{WhatIWant}) are either diagonal in the spin basis and sensitive only to $s_l$ mod $1$, or diagonal in the vortex basis (dual to the basis of spin-$1/2$ loops),
this is sufficient to prove that their critical behavior is identical.

We will carry out the proof in two steps.  First, we will show the equality for the two solvable points $\Gamma =0, K>0$ and $K=0, \Gamma>0$, where we can construct exactly the ground-states in both models.  We will then use perturbation theory to argue that the result holds throughout the phase diagram.

\subsection{Equal weighting of loops in the ground states at the solvable points}

For $K=0, \Gamma>0$, the ground state has $\sigma^x_l \equiv 1$ in the $\mathbb{Z}_2$ gauge theory, and $\sigma_l \in \{0,1 \}$ for the Levin-Wen model.  In this limit, for any $C$ we have trivially that $\langle \alpha_{1/2}(C) \rangle =\langle \alpha_{x}(C) \rangle =0$ and the result holds.

Focusing on the opposite limit ( $\Gamma =0, K>0$), let us construct the exact ground states in the two models.
We begin with the $\mathbb{Z}_2$ gauge theory.  Let $| 0 \rangle$ denote the state with $\sigma^x =1, \tau_s^x=1$ on all links and sites. This satisfies the Gauss law, but is not an eigenstate of the plaquette projector. To construct such an eigenstate, we take
\be \label{PTC}
|\Psi_{TC} \rangle =\frac{1}{\sqrt{N}}  \sum_{ n=1 }^{N_P } \sum_{  *\mathbb{P}_n } \prod_{P \in *\mathbb{P}_n } B_P | 0 \rangle =\frac{2}{\sqrt{N}}\sum_{\{C \}} |\Psi_{C} \rangle
\ee
where $ *\mathbb{P}_n$ runs over all possible distinct choices of $n$ plaquettes on the lattice, and $N$ is a normalization.   This sum generates all possible configurations $C$ of loops with $\sigma^x =-1$, weighted equally (each configuration is in fact generated twice, since $\prod_P B_P =1$).
%: if $\{C_1, ... C_N \}$ is a set of closed loops, then
%$ \prod_{B_P \in \text{Int}(C_1) }  ... \prod_{B_P \in \text{Int}(C_n) } |0 \rangle $
%is a state with $\sigma^x = -1$ on exactly this set of loops.  (Here $\text{Int}(C_i)$ denotes all plaquettes in the interior of the loop $C_i$.)
Since $B_P^2 = 1$, $ B_{P_i} \prod_{P \in *\mathbb{P}_n } B_P  = \prod_{P \in *\mathbb{P}_n' } B_P$, where $*\mathbb{P}_n'$ is $*\mathbb{P}_n$ with $P_i$ either added (if it was not originally in the set) or deleted (if it was).
It follows that
\begin{align*}
\frac{1}{\sqrt{N}}   \sum_{ n=1 }^{N_P } \sum_{  *\mathbb{P}_n } B_{P_i} \prod_{P \in *\mathbb{P}_n } B_P | 0 \rangle  =
\frac{1}{\sqrt{N}}  \sum_{ n=1 }^{N_P } \sum_{  *\mathbb{P}_n }  \prod_{P \in *\mathbb{P}_n } B_P | 0 \rangle
\end{align*}
and $|\Psi_{TC} \rangle$ is a ground state.

A similar construction can be used in the Levin-Wen models.  Let $|\Psi^e \rangle $ be a state satisfying (\ref{WhatIWant}), with $\sigma_l$ an integer for every link $l$ (careful inspection of these two conditions reveals that  $|\Psi^e \rangle$ is a superposition of configurations of closed spin-$1$ loops).  Now consider:
\be \label{PLW}
|\Psi_{LW} \rangle =\frac{1}{\sqrt{N}} \sum_{ n=1 }^{ N_P  } \sum_{  *\mathbb{P}_n } \prod_{P \in *\mathbb{P}_n } \frac{1}{\sqrt{2} } B_P^{1/2} | \Psi^e \rangle
\ee
We will show presently that
\be \label{WhatINeed}
\left( B_P^{1/2} \right )^2 =1 + B_P^1 \ \ \ \ B_P^{1/2} B_P^1 = B_P^1 B_P^{1/2} = B_P^{1/2}
\ee
  Using this fact, we have
\begin{align*}
 B_{P_i}^{1/2}  \prod_{P \in *\mathbb{P}_n } B_P^{1/2}  = \begin{cases} \prod_{P \in *\mathbb{P}_n\neg P_i } B_P^{1/2}  \left( \mathbf{1} + B_{P_i}^1  \right )  & P_i \in  *\mathbb{P}_n \\
\prod_{P \in *\mathbb{P}_n\cup P_i } B_P^{1/2}  & P_i \not\in  *\mathbb{P}_n
\end{cases}
\end{align*}
 It follows that
\be
 \frac{1}{\sqrt{2}} B_{P_i}^{1/2} |\Psi_{LW}  \rangle= |\Psi_{LW}\rangle
\ee
and $|\Psi_{LW} \rangle$ is a ground state.

Now, for any closed loop $L$, we can generate a configuration with $\sigma_l = 1/2$ on all links in $L$, and no other links, by acting on $| \Psi^e \rangle$ with the product of $B_P^{1/2}$ on all plaquettes inside the loop or all plaquettes outside the loop (these are the only such configurations on the right-hand side of Eq. (\ref{PLW})).
In the $\mathbb{Z}_2$ gauge theory the same holds for closed loops of $\sigma^x =-1$.   Hence given $C$, we have
\ba \label{alphaExpects}
\langle \alpha_{1/2}(C) \rangle% = \langle \Psi_{LW} | \prod_{l \in C } \delta_{\sigma_l, 1/2} |\Psi_{LW} \rangle \n
&=& \frac{1}{N} \langle \Psi^e | \left( \prod_{P \in *\mathbb{P}(C) }  \frac{1}{\sqrt{2}} B_P^{1/2} \right)^2 |\Psi^e \rangle \n
&=& \frac{1}{N} \langle \Psi^e |  \prod_{P \in *\mathbb{P}(C) }  \frac{1}{2}  \left( \mathbf{1} + B_P^1 \right ) |\Psi^e \rangle \n
&=& \frac{1}{N}
\ea
where $*\mathbb{P}(C)$ contains either all plaquettes inside, or all plaquettes outside, the closed loops in configuration $C$, and the last equality is a result of imposing (\ref{WhatIWant}).
Thus for $\Gamma =0$ all possible configurations of spin-$1/2$ loops occur with equal probability in the ground state of the SU(2)$_2$ Levin-Wen model.

We note that these results carry over directly to the more general case of an SU(2)$_k$ Levin-Wen model, upon replacing spin-$1/2$ (spin-$1$) with the set of all half-integer (integer)  spins, and $B_P^{1/2}$ ($B_P^1$) with the sum of all half-integer (integer) spin-raising terms in the plaquette operator (weighted by their respective quantum dimensions).

It remains to show that Eq. (\ref{WhatINeed}) holds, which we will do for general $k$.
We let $\mc{P}_P^{1/2} = \frac{1}{\mc{D}} \sum_{\sigma = 1/2, 3/2, ...} a_\sigma B_P^\sigma$ denote all half-integer raising terms in the plaquette operator, and $\mc{P}_P^1 = \frac{1}{\mc{D}}\left( \mathbf{1} +  \sum_{\sigma = 1, 2, ...} a_\sigma B_P^\sigma \right ) $ denote all integer terms.   Following Levin and Wen, we choose the constant $\mc{D}$ such that $\mc{P}_P \equiv \frac{1}{2} \left( \mc{P}_P^1 + \mc{P}_P^{1/2} \right ) $ is a projector.  We then have
\begin{align*}
(\mc{P}_P^1 + \mc{P}_P^{1/2})^2& = ( \mc{P}_P^1)^2 +   ( \mc{P}_P^{1/2})^2 + \mc{P}_P^{1/2} \mc{P}_P^1 + \mc{P}_P^1 \mc{P}_P^{1/2} \n
& = 2( \mc{P}_P^1 + \mc{P}_P^{1/2})
\end{align*}
Now, $ ( \mc{P}_P^{1/2})^2$ and $ ( \mc{P}_P^1)^2$ both contain only terms that raise the spins in $P$ by an integer amount, while $\mc{P}_P^1 \mc{P}_P^{1/2} = \mc{P}_P^{1/2} \mc{P}_P^1$ contains only half-integral raising operators.  It follows that
\be
 ( \mc{P}_P^1)^2 +   ( \mc{P}_P^{1/2})^2 =2  \mc{P}_P^1\,, \qquad \mc{P}_P^1 \mc{P}_P^{1/2} = \mc{P}_P^{1/2}.
 \ee
We also have
\be
( \mc{P}_P^1 + \mc{P}_P^{1/2}) (\mc{P}_P^1 - \mc{P}_P^{1/2}) =0
\ee
since it can be shown\cite{Burnell:2011kx} that $( \mc{P}_P^1 + \mc{P}_P^{1/2}) $ projects onto flux-free states, while $(\mc{P}_P^1 - \mc{P}_P^{1/2})$ projects onto states with an Ising vortex.  It follows that
\be \label{BP2}
(\mc{P}_P^{1/2})^2 = (\mc{P}_P^1)^2=  \mc{P}_P^1
\ee
This also implies that $ \mc{P}_P^1$ is a projector, and thus that the eigenvalues of $\mc{P}_P^1$ are $0$ and $1$.  (From Eq (\ref{BP2}) and the fact that $\mc{P}_P$ is a projector, it follows that the eigenvalues of $\mc{P}_P^{1/2}$ are $0, \pm 1$; when restricted to configurations where $\mc{P}_P^1 |\Psi \rangle = |\Psi \rangle$, they are $\pm 1$, as one expects from the correspondence of $\mc{P}_P^{1/2}$ to the plaquette term of the toric code.)

\subsection{Away from the solvable points}

Next, we wish to show that the result of the previous section holds true throughout the phase diagram.
One way to do this is to invoke the result of Ref. \onlinecite{Burnell:2011kx}, where it was shown that within the subspace of states satisfying (\ref{WhatIWant}), the SU(2)$_k$ Levin-Wen models are exactly dual to the transverse-field Ising model.  We can identify all states in this Hilbert space by the configuration of dual Ising spins (together with their topological ground-state sector, in periodic boundary conditions).   The duality relation--- which also holds for the $\mathbb{Z}_2$ gauge theory --- ensures that the probability amplitude to be in a given vortex configuration is identical in both models.  The physical operators in this Ising subspace are either diagonal in the vortex (or dual Ising spin) basis, or diagonal in the basis of spin-$1/2$ loops.   (These are precisely the operators that do not cause violations of (\ref{WhatIWant}), and cannot distinguish between edges of spin $0$ and spin $1$).   It follows that all expectation values of such operators --- including $\langle  \alpha_{1/2}(C) \rangle$--- must also be identical to their $\mathbb{Z}_2$  analogues (such as $\langle  \alpha_{x}(C) \rangle$).

Here we will take an alternative, perturbative approach to prove the desired result.
We will begin at an arbitrary point in the deconfined phase, and consider constructing the wave-functions in both theories to some finite order in perturbation theory.  These wave functions are linear combinations of the unperturbed (Levin-Wen or toric code) ground state, together with excited states of the form
\be
|\Psi_{ \{l \} } \rangle = \prod_{l \in \{l \} } h_l |\Psi_0 \rangle
\ee
where we have defined the transverse field operator $h_l \equiv \sigma^x_l$ for the toric code, and $(-1)^{2 s_l}$ for the Levin-Wen model, and $|\Psi_0 \rangle$ denotes the unperturbed ground state.
 If $l_1$ and $l_2$ are two links bordering plaquette $P$, we have
 \begin{align*}
\sigma^x_{l_{1,2} } B_P &= - B_P \sigma^x_{l_{1,2} } \\
(-1)^{2 s_{l_{1,2} } } B^{1/2}_P& = - B^{1/2}_P(-1)^{2 s_{l_{1,2} } }  \\
 \left[ \sigma^x_{l_{1} } \sigma^x_{l_{2} } , B_P \right ] &= \left [(-1)^{2 s_{l_{1} } }(-1)^{2 s_{l_{2} } } , B^{1/2}_P \right  ] =0
 \end{align*}
Thus $|\Psi_{ \{l \} } \rangle$ is a state with vortices on each plaquette with an odd number of edges in the set of links $\{l \}$.  It also follows that choices of $\{l \}$ which differ by a product  $\prod_{l \in C^* } h_l$, where $C^*$ is a set of closed curves on the dual lattice, create identical excited states, as  $\prod_{l \in C^* } h_l |\Psi_0 \rangle =| \Psi_0 \rangle$.
Finally, we have
\be
\langle \Psi_{ \{l \} }  | \Psi_{ \{l' \} }  \rangle = \delta_{ \{l\} \cup \{ l' \}, C^* }
\ee
In other words, the inner product is $1$ if the combination of the two sets $ \{ l \}$ and $\{ l' \}$ of links forms a set of closed curves on the dual lattice, so that $ | \Psi_{ \{l \} }  \rangle$ and $ | \Psi_{ \{l' \} }  \rangle$ have vortices on the same plaquettes.
Similarly, we may compute matrix elements of the Hamiltonian within these excited states via:
\be \label{MyLls}
\langle \Psi_{ \{l \} }  | \prod_{l \in \{ l'' \} } h_l | \Psi_{ \{l' \} }  \rangle = \delta_{ \{l\} \cup \{ l' \} \cup \{ l'' \} , C^* }
\ee

The crucial point is that for any choice of $\{l\} ,\{ l' \},\{ l'' \}$, these matrix elements are identical in both models.
Since the weight of each unperturbed excited state in the exact ground state can be constructed perturbatively using only matrix elements of this form, it follows that
\be
|\Psi \rangle = \sum_{ \{ l \} } \beta^\Gamma_{\{ l \} }  | \Psi_{ \{l \} }  \rangle
\ee
with $ \beta_{\{ l \} }$ the same for each set $\{l \}$ in both models.

Finally, we observe that $ \alpha_{1/2}(C)$ and $\alpha_x(C)$ both have the form
\be
\alpha_{\nu} = \prod_{l \in C} \frac{1}{2} \left( 1 - h_l \right)
\ee
and, in particular, commute with $h_l$ on every link.   (Here $\alpha_\nu =\alpha_{1/2}(C), \alpha_{x}(C)$ as appropriate).
This, together with the relation (\ref{MyLls}), implies that
\begin{align*}
\langle\alpha_{\nu}  \rangle_{\Gamma}  &=  \sum_{  \{l \} ,  \{l' \} }  \overline{\beta}^\Gamma_{\{l \} } \ \beta^\Gamma_ {\{l' \} } \langle \Psi_{0} |  \prod_{l \in \{l \} } h_l  \    \alpha_{\nu}  \prod_{l \in \{l' \} } h_l   |\Psi_{0} \rangle 
%&=  \sum_{  \{l \} ,  \{l' \} } \delta_{ \{l\} \cup \{ l' \} , C^* } \overline{\beta}^\Gamma _{\{l \} }  \beta^\Gamma_ {\{l' \} } \langle \Psi_{LW}^{(0)} |   \alpha_{\nu}  |\Psi_{LW}^{(0)} \rangle
\end{align*}
in both models.
We have already shown that the coefficients $\beta^\Gamma_{ \{l \}}$ are the same, and the possible choices of $\{l \} , \{ l' \}$ on which the $\delta$ function has support are a geometric property of the lattice.  Invoking the result of the previous subsection, we can thus conclude that for all $\Gamma$ in the deconfined phase,
\be \label{ExpGamma}
\langle \alpha_{1/2}(C)  \rangle_\Gamma = \langle \alpha_x(C) \rangle_\Gamma
\ee

Our derivation has implicitly relied on the fact that we can construct the exact ground state perturbatively, starting from the ground state of the toric code or Levin-Wen solvable point.  Thus the above argument fails at the critical point, and in the phase where $\Gamma/K$ is large.  In this regime, however, we may make essentially the same argument, by replacing $h_l$ with the plaquette operator, and $|\Psi_{LW} \rangle, |\Psi_{TC} \rangle$ (denoted by $|\Psi_0\rangle$ in the derivation above) with $|\Psi^e \rangle$ and $|0 \rangle$ respectively.  In this case, the basis of excited states generated will be an eigenstate of $\sigma^x$ (toric code) or $(-1)^{2 s}$ (Levin-Wen).   There is no need to define an analogue of $C^*$, since if two distinct products of plaquette projectors produce the same loop configuration state, then their product is the identity operator.

In each phase, we can thus argue that Eq. (\ref{ExpGamma}) holds to arbitrary order in perturbation theory.  It follows that as the phase transition is second order, it must also hold at the critical point, proving the result.

\end{appendix}

\bibliography{main}

\end{document}